\documentclass[11pt]{article}
\usepackage{hyperref,url}
\usepackage[utf8x]{inputenc} 
\usepackage{latexsym,graphicx,mathrsfs,amsfonts,subfig}
\usepackage{slashed, color}
\usepackage{multirow}
\usepackage{tikz}
\usepackage{empheq}
\usepackage{amsmath}
\allowdisplaybreaks[2]
\usepackage[multiple]{footmisc}%for multiple footnotes
\usepackage{textcomp}
\usepackage{amssymb}%for semidirect product symbol
\usetikzlibrary{positioning}
\usepackage{amscd}
\usepackage{amsthm}
\usepackage{array} 

\renewcommand{\thefootnote}{\fnsymbol{footnote}}
\numberwithin{equation}{section}
\usepackage{dsfont}
\usepackage[toc,page]{appendix}

\DeclareFontFamily{U}{MnSymbolC}{}
\DeclareSymbolFont{MnSyC}{U}{MnSymbolC}{m}{n}
\DeclareFontShape{U}{MnSymbolC}{m}{n}{
	<-6>  MnSymbolC5
	<6-7>  MnSymbolC6
	<7-8>  MnSymbolC7
	<8-9>  MnSymbolC8
	<9-10> MnSymbolC9
	<10-12> MnSymbolC10
	<12->   MnSymbolC12}{}
\DeclareMathSymbol{\intprod}{\mathbin}{MnSyC}{'270}
\newcommand{\ov}{\overline}

\newcommand{\til}{\widetilde}

\let\nc\newcommand
\let\renc\renewcommand
\nc{\wbar}{\overline}
\let\td\tilde
\let\wtd\widetilde
\let\wht\widehat
\let\mcl\mathcal

\nc{\ab}{{\bar{a}}} \nc{\at}{\tilde{a}} \nc{\ah}{\hat{a}}
\nc{\bb}{{\bar{b}}} 
\nc{\bh}{\hat{b}}
\nc{\cb}{{\bar{c}}} \nc{\ct}{\tilde{c}} %\nc{\ch}{\hat{c}}
\nc{\db}{{\bar{d}}} \nc{\dt}{\tilde{d}} \renc{\dh}{\hat{d}}
\nc{\eb}{{\bar{e}}} \nc{\et}{\tilde{e}} \nc{\eh}{\hat{e}}
\nc{\fb}{{\bar{f}}} \nc{\ft}{\tilde{f}} \nc{\fh}{\hat{f}}
\nc{\ib}{{\bar{\imath}}} \nc{\ih}{\hat{\imath}} %\nc{\it}{\tilde{\imath}}
\nc{\jb}{{\bar{\jmath}}} \nc{\jt}{\tilde{\jmath}} \nc{\jh}{\hat{\jmath}}
\nc{\kb}{{\bar{k}}} \nc{\kt}{\tilde{k}} \nc{\kh}{\hat{k}}
\nc{\lb}{{\bar{l}}} \nc{\lt}{\tilde{l}} \nc{\lh}{\hat{l}}
\nc{\mb}{{\bar{m}}} \nc{\mt}{\tilde{m}} \nc{\mh}{\hat{m}}
\nc{\nb}{{\bar{n}}} \nc{\nt}{\tilde{n}} \nc{\nh}{\hat{n}}
\nc{\ob}{{\bar{o}}} \nc{\ot}{\tilde{o}} \nc{\oh}{\hat{o}}
\nc{\pb}{{\bar{p}}} \nc{\pt}{\tilde{p}} \nc{\ph}{\hat{p}}
\nc{\qb}{{\bar{q}}} \nc{\qt}{\tilde{q}} \nc{\qh}{\hat{q}}
\nc{\rb}{{\bar{r}}} \nc{\rt}{\tilde{r}} \nc{\rh}{\hat{r}}
\renc{\sb}{{\bar{s}}} \nc{\st}{\tilde{s}} \nc{\sh}{\hat{s}}
\nc{\tb}{{\bar{t}}} \renc{\th}{\hat{t}} %\nc{\tt}{\tilde{t}}
\nc{\ub}{{\bar{u}}} \nc{\ut}{\tilde{u}} \nc{\uh}{\hat{u}}
\nc{\vb}{{\bar{v}}} \nc{\vt}{\tilde{v}} \nc{\vh}{\hat{v}}
\nc{\wt}{\tilde{w}} \nc{\wh}{\hat{w}}
\nc{\xb}{{\bar{x}}} \nc{\xt}{\tilde{x}} \nc{\xh}{\hat{x}}
\nc{\yb}{{\bar{y}}} \nc{\yt}{\tilde{y}} \nc{\yh}{\hat{y}}
\nc{\zb}{{\bar{z}}} \nc{\zt}{\tilde{z}} 
\nc{\Ab}{\wbar{A}} \nc{\At}{\wtd{A}} \nc{\Ah}{\wht{A}}
\nc{\Bb}{\wbar{B}} \nc{\Bt}{\wtd{B}} \nc{\Bh}{\wht{B}}
\nc{\Cb}{\wbar{C}} \nc{\Ct}{\wtd{C}} \nc{\Ch}{\wht{C}}
\nc{\Db}{\wbar{D}} \nc{\Dt}{\wtd{D}} \nc{\Dh}{\wht{D}}
\nc{\Eb}{\wbar{E}} \nc{\Et}{\wtd{E}} \nc{\Eh}{\wht{E}}
\nc{\Fb}{\wbar{F}} \nc{\Ft}{\wtd{F}} \nc{\Fh}{\wht{F}}
\nc{\Gb}{\wbar{G}} \nc{\Gt}{\wtd{G}} \nc{\Gh}{\wht{G}}
\nc{\Hb}{\wbar{H}} \nc{\Ht}{\wtd{H}} \nc{\Hh}{\wht{H}}
\nc{\Ib}{\wbar{I}} \nc{\It}{\wtd{I}} \nc{\Ih}{\wht{I}}
\nc{\Jb}{\wbar{J}} \nc{\Jt}{\wtd{J}} \nc{\Jh}{\wht{J}}
\nc{\Kb}{\wbar{K}} \nc{\Kt}{\wtd{K}} \nc{\Kh}{\wht{K}}
\nc{\Lb}{\wbar{L}} \nc{\Lt}{\wtd{L}} \nc{\Lh}{\wht{L}}
\nc{\Mb}{\wbar{M}} \nc{\Mt}{\wtd{M}} \nc{\Mh}{\wht{M}}
\nc{\Nb}{\wbar{N}} \nc{\Nt}{\wtd{N}} \nc{\Nh}{\wht{N}}
\nc{\Ob}{\wbar{O}} \nc{\Ot}{\wtd{O}} \nc{\Oh}{\wht{O}}
\nc{\Pb}{\wbar{P}} \nc{\Pt}{\wtd{P}} \nc{\Ph}{\wht{P}}
\nc{\Qb}{\wbar{Q}} \nc{\Qt}{\wtd{Q}} \nc{\Qh}{\wht{Q}}
\nc{\Rb}{\wbar{R}} \nc{\Rt}{\wtd{R}} \nc{\Rh}{\wht{R}}
\nc{\Sb}{\wbar{S}} \nc{\St}{\wtd{S}} \nc{\Sh}{\wht{S}}
\nc{\Tb}{\wbar{T}} \nc{\Tt}{\wtd{T}} \nc{\Th}{\wht{T}}
\nc{\Ub}{\wbar{U}} \nc{\Ut}{\wtd{U}} \nc{\Uh}{\wht{U}}
\nc{\Vb}{\wbar{V}} \nc{\Vt}{\wtd{V}} \nc{\Vh}{\wht{V}}
\nc{\Wb}{\wbar{W}} \nc{\Wt}{\wtd{W}} \nc{\Wh}{\wht{W}}
\nc{\Xb}{\wbar{X}} \nc{\Xt}{\wtd{X}} \nc{\Xh}{\wht{X}}
\nc{\Yb}{\wbar{Y}} \nc{\Yt}{\wtd{Y}} \nc{\Yh}{\wht{Y}}
\nc{\Zb}{\wbar{Z}} \nc{\Zt}{\wtd{Z}} \nc{\Zh}{\wht{Z}}

\nc{\CA}{\mcl{A}} \nc{\CAb}{\wbar{\CA}} \nc{\CAt}{\wtd{\CA}} \nc{\CAh}{\wht{\CA}}
\nc{\CB}{\mcl{B}} \nc{\CBb}{\wbar{\CB}} \nc{\CBt}{\wtd{\CB}} \nc{\CBh}{\wht{\CB}}
\nc{\CC}{\mcl{C}} \nc{\CCb}{\wbar{\CC}} \nc{\CCt}{\wtd{\CC}} \nc{\CCh}{\wht{\CC}}
\nc{\cDt}{\wtd{\cC}} \nc{\cDh}{\wht{\cD}}
\nc{\CE}{\mcl{E}} \nc{\CEb}{\wbar{\CE}} \nc{\CEt}{\wtd{\CE}} \nc{\CEh}{\wht{\CE}}
\nc{\CF}{\mcl{F}} \nc{\CFb}{\wbar{\CF}} \nc{\CFt}{\wtd{\CF}} \nc{\CFh}{\wht{\CF}}
\nc{\CG}{\mcl{G}} \nc{\CGb}{\wbar{\CG}} \nc{\CGt}{\wtd{\CG}} \nc{\CGh}{\wht{\CG}}
\nc{\CH}{\mcl{H}} \nc{\CHb}{\wbar{\CH}} \nc{\CHt}{\wtd{\CH}} \nc{\CHh}{\wht{\CH}}
\nc{\CI}{\mcl{I}} \nc{\CIb}{\wbar{\CI}} \nc{\CIt}{\wtd{\CI}} \nc{\CIh}{\wht{\CI}}
\nc{\CJ}{\mcl{J}} \nc{\CJb}{\wbar{\CJ}} \nc{\CJt}{\wtd{\CJ}} \nc{\CJh}{\wht{\CJ}}
\nc{\CK}{\mcl{K}} \nc{\CKb}{\wbar{\CK}} \nc{\CKt}{\wtd{\CK}} \nc{\CKh}{\wht{\CK}}
\nc{\CL}{\mcl{L}} \nc{\CLb}{\wbar{\CL}} \nc{\CLt}{\wtd{\CL}} \nc{\CLh}{\wht{\CL}}
\nc{\CM}{\mcl{M}} \nc{\CMb}{\wbar{\CM}} \nc{\CMt}{\wtd{\CM}} \nc{\CMh}{\wht{\CM}}
\nc{\CN}{\mcl{N}} \nc{\CNb}{\wbar{\CN}} \nc{\CNt}{\wtd{\CN}} \nc{\CNh}{\wht{\CN}}
\nc{\CO}{\mcl{O}} \nc{\COb}{\wbar{\CO}} \nc{\COt}{\wtd{\CO}} \nc{\COh}{\wht{\CO}}
\nc{\CQ}{\mcl{Q}} \nc{\CQb}{\wbar{\CQ}} \nc{\CQt}{\wtd{\CQ}} \nc{\CQh}{\wht{\CQ}}
\nc{\CR}{\mcl{R}} \nc{\CRb}{\wbar{\CR}} \nc{\CRt}{\wtd{\CR}} \nc{\CRh}{\wht{\CR}}
\nc{\CS}{\mcl{S}} \nc{\CSb}{\wbar{\CS}} \nc{\CSt}{\wtd{\CS}} \nc{\CSh}{\wht{\CS}}
\nc{\CT}{\mcl{T}} \nc{\CTb}{\wbar{\CT}} \nc{\CTt}{\wtd{\CT}} \nc{\CTh}{\wht{\CT}}
\nc{\CU}{\mcl{U}} \nc{\CUb}{\wbar{\CU}} \nc{\CUt}{\wtd{\CU}} \nc{\CUh}{\wht{\CU}}
\nc{\CV}{\mcl{V}} \nc{\CVb}{\wbar{\CV}} \nc{\CVt}{\wtd{\CV}} \nc{\CVh}{\wht{\CV}}
\nc{\CW}{\mcl{W}} \nc{\CWb}{\wbar{\CW}} \nc{\CWt}{\wtd{\CW}} \nc{\CWh}{\wht{\CW}}
\nc{\CX}{\mcl{X}} \nc{\CXb}{\wbar{\CX}} \nc{\CXt}{\wtd{\CX}} \nc{\CXh}{\wht{\CX}}
\nc{\CY}{\mcl{Y}} \nc{\CYb}{\wbar{\CY}} \nc{\CYt}{\wtd{\CY}} \nc{\CYh}{\wht{\CY}}
\nc{\CZ}{\mcl{Z}} \nc{\CZb}{\wbar{\CZ}} \nc{\CZt}{\wtd{\CZ}} \nc{\CZh}{\wht{\CZ}}

\let\eps\epsilon
\let\ups\upsilon
\let\veps\varepsilon
\let\vtht\vartheta
\let\vsgm\varsigma
\let\vphi\varphi
\let\vrho\varrho

\nc{\alphab}{\bar{\alpha}} \nc{\alphat}{\td{\alpha}} \nc{\alphah}{\hat{\alpha}}
\nc{\betab}{\bar{\beta}}   \nc{\betat}{\td{\beta}}   \nc{\betah}{\hat{\beta}} 
\nc{\gammab}{\bar{\gamma}} \nc{\gammat}{\td{\gamma}} \nc{\gammah}{\hat{\gamma}} 
\nc{\deltab}{\bar{\delta}} \nc{\deltat}{\td{\delta}} \nc{\deltah}{\hat{\delta}} 
\nc{\epsilonb}{\bar{\eps}} \nc{\epsilont}{\td{\eps}} \nc{\epsilonh}{\hat{\eps}} 
\nc{\vepsb}{\bar{\veps}}   \nc{\vepst}{\td{\veps}}   \nc{\vepsh}{\hat{\veps}} 
\nc{\zetab}{\bar{\zeta}}   \nc{\zetat}{\td{\zeta}}   \nc{\zetah}{\hat{\zeta}} 
\nc{\etab}{\bar{\eta}}     
\nc{\etah}{\hat{\eta}} 
\nc{\thetab}{\bar{\theta}} \nc{\thetat}{\td{\theta}} \nc{\thetah}{\hat{\theta}} 
\nc{\vthetab}{\bar{\vtht}} \nc{\vthetat}{\td{\vtht}} \nc{\vthetah}{\hat{\vtht}} 
\nc{\lambdat}{\td{\lambda}} \nc{\lambdah}{\hat{\lambda}} 
\nc{\iotab}{\bar{\iota}}   \nc{\iotat}{\td{\iota}}   \nc{\iotah}{\hat{\iota}} 
\nc{\kappab}{\bar{\kappa}} \nc{\kappat}{\td{\kappa}} \nc{\kappah}{\hat{\kappa}} 
\nc{\lmdb}{\bar{\lmd}}     \nc{\lmdt}{\td{\lmd}}     \nc{\lmdh}{\hat{\lmd}} 
\nc{\mub}{\bar{\mu}}       \nc{\mut}{\td{\mu}}       \nc{\muh}{\hat{\mu}} 
\nc{\nub}{\bar{\nu}}       \nc{\nut}{\td{\nu}}       \nc{\nuh}{\hat{\nu}} 
\nc{\xib}{\bar{\xi}}       \nc{\xit}{\td{\xi}}       \nc{\xih}{\hat{\xi}} 
\nc{\pib}{\bar{\pi}}       \nc{\pit}{\td{\pi}}       \nc{\pih}{\hat{\pi}} 
\nc{\vpib}{\bar{\vpi}}     \nc{\vpit}{\td{\vpi}}     \nc{\vpih}{\hat{\vpi}} 
\nc{\rhob}{\bar{\rho}}     \nc{\rhot}{\td{\rho}}     \nc{\rhoh}{\hat{\rho}} 
\nc{\vrhob}{\bar{\vrho}}   \nc{\vrhot}{\td{\vrho}}   \nc{\vrhoh}{\hat{\vrho}} 
\nc{\sigmab}{\bar{\sigma}} \nc{\sigmat}{\td{\sigma}} \nc{\sigmah}{\hat{\sigma}} 
\nc{\vsigmab}{\bar{\vsgm}} \nc{\vsigmat}{\td{\vsgm}} \nc{\vsigmah}{\hat{\vsgm}} 
\nc{\taub}{\bar{\tau}}     \nc{\taut}{\td{\tau}}     \nc{\tauh}{\hat{\tau}} 
\nc{\upsb}{\bar{\ups}} \nc{\upst}{\td{\ups}} \nc{\upsh}{\hat{\ups}} 
\nc{\phib}{\bar{\phi}}     \nc{\phit}{\td{\phi}}     \nc{\phih}{\hat{\phi}} 
\nc{\varphib}{\bar{\vphi}}   \nc{\varphit}{\td{\vphi}}   \nc{\varphih}{\hat{\vphi}} 
\nc{\chib}{\bar{\chi}}     
\nc{\chih}{\hat{\chi}} 
\nc{\psib}{\bar{\psi}}     
\nc{\psih}{\hat{\psi}} 
\nc{\omegab}{\bar{\omega}} \nc{\omegat}{\td{\omega}} \nc{\omegah}{\hat{\omega}} 

\nc{\Gammab}{\wbar{\Gamma}}     \nc{\Gammat}{\wtd{\Gamma}}     \nc{\Gammah}{\wht{\Gamma}}
\nc{\Deltab}{\wbar{\Delta}}     \nc{\Deltat}{\wtd{\Delta}}     \nc{\Deltah}{\wht{\Delta}}
\nc{\Thetab}{\wbar{\Theta}}     \nc{\Thetat}{\wtd{\Theta}}     \nc{\Thetah}{\wht{\Theta}}
\nc{\Lambdab}{\wbar{\Lambda}}   \nc{\Lambdat}{\wtd{\Lambda}}   \nc{\Lambdah}{\wht{\Lambda}}
\nc{\Xib}{\wbar{\Xi}}           \nc{\Xit}{\wtd{\Xi}}           \nc{\Xih}{\wht{\Xi}}
\nc{\Pib}{\wbar{\Pi}}           \nc{\Pit}{\wtd{\Pi}}           \nc{\Pih}{\wht{\Pi}}
\nc{\Sigmab}{\wbar{\Sigma}}     \nc{\Sigmat}{\wtd{\Sigma}}     \nc{\Sigmah}{\wht{\Sigma}}
\nc{\Upsilonb}{\wbar{\Upsilon}} \nc{\Upsilont}{\wtd{\Upsilon}} \nc{\Upsilonh}{\wht{\Upsilon}}
\nc{\Phib}{\wbar{\Phi}}         \nc{\Phit}{\wtd{\Phi}}         \nc{\Phih}{\wht{\Phi}}
\nc{\Psib}{\wbar{\Psi}}         \nc{\Psit}{\wtd{\Psi}}         \nc{\Psih}{\wht{\Psi}}
\nc{\Omegab}{\wbar{\Omega}}     \nc{\Omegat}{\wtd{\Omega}}     \nc{\Omegah}{\wht{\Omega}}
\nc{\Varepsilon}{\mathcal{E}}

\newcommand{\cD}{{\cal D}}

\nc{\balpha}{\bar{\alpha}}
\nc{\bbeta}{\bar{\beta}}
\nc{\bgamma}{\bar{\gamma}}
\nc{\bm}{\bar{m}}
\nc{\bn}{\bar{n}}
\nc{\bp}{\bar{p}}
\nc{\al}{\alpha}
\nc{\bt}{\beta}
\nc{\gm}{\gamma}
\nc{\zh}{\wht{z}}
\nc{\zhb}{\ov{\wht{z}}}
\nc{\mbh}{\wht{\ov{m}}}
\nc{\bc}{|_{x^2=0}}

\nc{\tal}{\til{\al}}
\nc{\tbt}{\til{\bt}}
\nc{\tgm}{\til{\gm}}

\nc{\wb}{\ov{w}}
\nc{\teta}{\til{\eta}}
\nc{\tpsi}{\til{\psi}}

\def\IL{\relax{\rm I\kern-.18em L}}
\def\IH{\relax{\rm I\kern-.18em H}}
\def\IB{\relax{\rm I\kern-.18em B}}
\def\ID{\relax{\rm I\kern-.18em D}}
\def\IE{\relax{\rm I\kern-.18em E}}
\def\IF{\relax{\rm I\kern-.18em F}}

\def\IG{\relax\hbox{$\inbar\kern-.3em{\rm G}$}}
\def\IGa{\relax\hbox{${\rm I}\kern-.18em\Gamma$}}
\def\IH{\relax{\rm I\kern-.18em H}}
\def\II{\relax{\rm I\kern-.18em I}}
\def\IK{\relax{\rm I\kern-.18em K}}
\def\IP{\relax{\rm I\kern-.18em P}}
\def\IQ{\relax\hbox{$\inbar\kern-.3em{\rm Q}$}}

\def\hat{\widehat}
\def\CM {{\cal M}}
\def\CN {{\cal N}}
\def\CR {{\cal R}}

\def\CF {{\cal F}}
\def\CJ {{\cal J}}

\def\CL {{\cal L}}
\def\CV {{\cal V}}
\def\CO {{\cal O}}
\def\CZ {{\cal Z}}
\def\CE {{\cal E}}
\def\CG {{\cal G}}
\def\CH {{\cal H}}
\def\CC {{\cal C}}
\def\CB {{\cal B}}
\def\CS {{\cal S}}
\def\CA{{\cal A}}
\def\CK{{\cal K}}
\def\CQ{{\cal Q}}

\def\p{\partial}
\def\pb{{\bar \p}}

\def\vt#1#2#3{ {\vartheta[{#1 \atop  #2}](#3\vert \tau)} }

\def\jb{{\bar j}}

\def\inbar{\,\vrule height1.5ex width.4pt depth0pt}

\nc{\hTheta}{\hat{\Theta}}
\nc{\vp}{\varphi}
\nc{\tg}{\widetilde{g}}

\let\OLDthebibliography\thebibliography
\renewcommand\thebibliography[1]{
	\OLDthebibliography{#1}
	\setlength{\parskip}{5pt}
	\setlength{\itemsep}{0pt plus 0.3ex}
}
\usepackage{titlesec}
\titleformat*{\section}{\bfseries\large}
\begin{document}
\addtolength{\baselineskip}{1.5mm}

\thispagestyle{empty}
%\begin{flushright}
%hep-th/   \\
%\end{flushright}
\vbox{}
\vspace{3.0cm}

\begin{center}
\centerline{\LARGE{   Implications of Tarski's Undefinability Theorem on the Theory of Everything}}

% \begin{center}
% 	\centerline{\LARGE{Noncommutative Toda Field Theories }}
% 	\bigskip
% \centerline{\LARGE{from Noncommutative 4d/5d/6d Chern-Simons Theories}}
 %\centerline{\LARGE{and non-BPS Domain Walls}} \\ 
	%\medskip
	
\vspace{2.5cm}
%\vspace{3.0cm}
		
	{{Mir Faizal$^{1,2,3}$, Arshid Shabir$^1$, Aatif Kaisar Khan$^{1,4}$}}
	\\{
\textit{\small $^{1}$Canadian Quantum Research Center, 204-3002, 32 Ave Vernon, BC V1T 2L7, Canada}\\
\textit{\small $^{2}$Irving K. Barber School of Arts and Sciences, University of British Columbia Okanagan, Kelowna, BC V1V 1V7, Canada}\\
\textit{\small $^{3}$Department of Mathematical Sciences, Durham University,
Lower Mountjoy, Stockton Road, Durham DH1 3LE, UK}\\
\textit{\small $^{4}$Facultat de Física, Universitat de Barcelona (UB), Martí i Franquès 1, E08028, Barcelona, Spain}
}
\end{center}

\vspace{2.0cm}

\centerline{\bf Abstract}\smallskip \noindent
\
The Theory of Everything ($S_{\text{ToE}}$) seeks to unify all fundamental forces of nature, including quantum gravity, into a single theoretical framework.  
This theory would be defined internally using a set of axioms, and this paper proposes a set of axioms for any such theory. 
Furthermore, for such a theory, all scientific truth would be defined internally as consequences derivable from the rules of such a theory. This paper then examines the implications of Tarski's undefinability theorem on scientific truths derived from such axioms. We demonstrate that  Tarski's theorem imposes limitations on any such formal system $S_{\text{ToE}}$. However, we also argue that the Lucas-Penrose argument suggests that non-algorithmic understanding can transcend these formal limitations. 
\newpage

\renewcommand{\thefootnote}{\arabic{footnote}}
\setcounter{footnote}{0}

%\tableofcontents
%\bigskip\noindent  \section{Introduction}
%\textcolor{blue}{The \(S_{\text{ToE}}\) ($S_{\text{ToE}}$) aims to create a unified framework encompassing all fundamental forces, including quantum gravity.[similar to the starting of the abstract]}  The goal is to reconcile quantum mechanics and general relativity, resolving the inconsistencies that emerge when merging these essential pillars of modern physics. 

The discovery of the Theory of Everything ($S_{\text{ToE}}$) can have important consequences for the very definition of scientific truths, and scientific methods. 
Scientific truth is usually associated with empirical evidence and reproducibility, emphasizing observations, experiments, and the formulation of theories that could be tested and potentially falsified~\cite{Galilei1967}. 
This empirical approach has been further developed, and the scientific method placed a strong emphasis on hypothesis testing,  and a rigorous process of validation. Thus, scientific truth became less about absolute certainty and more about the best possible explanation given the available evidence. This perspective is epitomized by Karl Popper's philosophy of science, which argues that scientific truths are provisional and should always be open to falsification~\cite{Popper1959}. However, if a theory is tested to be correct, then its consequences are also viewed as scientific truths (even if they cannot be directly tested). For example, as general relativity has been tested to be a correct theory of nature, then all its consequences (like black holes, the physics within black holes, etc.) also become scientific. 
Now a theory can be falsified or be an approximation to a more correct theory, and in this case, some consequences of the theory might not be correct. However, if we do have a $S_{\text{ToE}}$ that represents all fundamental physical aspects of the universe/multiverse (even giving rise to the universe/multiverse), then scientific truth would be consequences derivable from such a theory.    So, truths within a $S_{\text{ToE}}$ should be internally consistent and self-contained within that framework.
In this scenario, scientific truths would be derived directly from the axioms  that constitute the $S_{\text{ToE}}$. This would imply a form of completeness in which all physical phenomena could be explained within a single, coherent theory. However, this also means that scientific truth would become internally defined, where the validity of a truth claim depends on its consistency with the $S_{\text{ToE}}$ rather than empirical verification alone~\cite{Weinberg1994}.

Quantum mechanics and general relativity represent the two core foundations of modern theoretical physics. Despite notable advances, fully unifying these theories into a comprehensive structure—where even spacetime is quantized—remains challenging. Although a consistent quantum theory of gravity is yet to be formulated, various approaches have surfaced, offering valuable predictions about quantum gravity's nature. A critical observation is that investigating spacetime at smaller scales requires increasingly higher energy levels. So,   probing the Planck scale requires Planck energy, but introducing Planck energy in that region would form a mini black hole, preventing measurements \cite{bl}. Since any theory of quantum gravity should be consistent with classical black hole physics, this implies quantum gravity includes a fundamental minimal length \cite{le} and minimal time \cite{ti}, potentially resolving issues in quantum field theory where calculations lead to divergences or infinities. Typically, a cutoff is introduced to limit the scale probed by the system, but if spacetime naturally includes a minimal length and time, this would inherently function as a cutoff \cite{cut}.

In perturbative string theory, a minimum length arises because spacetime cannot be examined below the string length scale \cite{pert}, the smallest possible probe size. According to quantum mechanics, an unmeasurable object doesn't physically exist, implying that spacetime beneath the string length doesn't exist. This principle holds even for point-like objects, such as D0-branes, in non-perturbative string theory due to the theory's dualities. T-duality, which connects string theory at large and small scales, shows that the spectrum above the string length scale is the same as the spectrum below it \cite{tdual}, confirming no new information is accessible beyond this minimal scale. {Other approaches to quantum gravity with discrete spacetime, such as causal sets \cite{c1}, quantum graphity \cite{c2}, and causal dynamical triangulation \cite{c4}, also predict minimal lengths, where geometry can't be probed below a certain scale.} This minimal geometric cutoff seems common across different approaches to quantum gravity,  and is  also found in noncommutative \cite{nc10} and nonlocal quantum field theories \cite{nl10}. {While minimal areas and volumes naturally arise in loop quantum gravity \cite{c6} and spin foam models \cite{c10},} loop quantum cosmology introduces a polymer length through a background-independent quantization method called polymer quantization \cite{c5}, This indicates that this is a general feature of quantum gravity \cite{c51}. Thus, $S_{\text{ToE}}$ cannot be a theory in spacetime, but rather a theory from which spacetime should emerge dynamically as an emergent phenomenon and breakdown at the Planck scale. 

This geometric cutoff helps address issues in general relativity, like singularities where spacetime models fail. The Penrose-Hawking theorems suggest that singularities are inherent to general relativity \cite{ph1, ph2}, but quantum gravity effects, with the geometric cutoff, modify these predictions, preventing singularity formation \cite{li1, li2}. In string theory, T-duality introduces a minimal length that also avoids singularities \cite{tdual1}, while loop quantum cosmology avoids singularities via a geometric cutoff \cite{lqgs}. This absence of singularities seems to be a universal feature of quantum gravity theories, as the Penrose-Hawking singularity theorems are connected to Bekenstein-Hawking entropy \cite{hpbe}, and a bound on this entropy, imposed by the geometric cutoff, prevents singularities.
Bekenstein-Hawking entropy is directly linked to geometry through the Jacobson formalism \cite{jc}, and modifying this entropy alters spacetime geometry. It has been shown that a bound on this entropy, derived from the minimal length in quantum gravity, prevents singularity formation in spacetime \cite{gup, gup2}. Singularities are avoided because a minimal value for a geometric quantity suggests spacetime geometry is an emergent structure that breaks down below this threshold. In general relativity, singularities arise when the theory is applied in contexts where spacetime descriptions become invalid. Interestingly, this geometric bound is closely tied to limits on quantum information, indicating that spacetime geometry itself may emerge from quantum information. The critical role played by the Bekenstein-Hawking entropy seems to indicate that information is more fundamental than spacetime, and spacetime emerges from the information of quantum states. 

The holographic principle best illustrates how spacetime geometry emerges from non-geometric quantum states. This principle connects a theory within a volume to one on its boundary \cite{holog}, widely applied in string theory, where it relates a gravitational theory in anti-de Sitter spacetime to a quantum field theory on its boundary \cite{ads}. It shows that the entanglement of quantum states in the boundary theory generates the geometric structure, and removing this entanglement destroys the geometry in the dual theory. 
Spacetime geometry only emerges at scales larger than the Planck scale, much like how a table's geometry arises from atomic physics at scales larger than atoms \cite{Faiz1}.
Various approaches to quantum gravity also suggest that geometry emerges from quantum information theory \cite{Faizal, Faiz1}. Thus, the emergence of spacetime geometry from information appears to be a model-independent feature of $S_{\text{ToE}}$.

Based on these insights, we propose a set of axioms for a $S_{\text{ToE}}$, including a theory of quantum spacetime (i.e., quantum gravity), applicable to any theory in this domain. These axioms highlight the role of information in the emergence of spacetime and suggest grounding quantum gravity within the framework of information theory.  As these axioms produce spacetime, they cannot be defined in spacetime\cite{Faizal, Faiz1}. These axioms, exist in a Platonic realm outside spacetime and reduce to quantum mechanics and general relativity in appropriate limits. Additionally, a computational algorithm in this Platonic realm actualizes the consequences of these axioms. We also examine Tarski's undefinability theorem \cite{t1, t2} within the context of $S_{\text{ToE}}$ and argue, using a generalization of the Lucas-Penrose argument \cite{penrose1989emperor, lucas1961minds}, that a non-algorithmic understanding, separate from any computational algorithm, must exist in this Platonic realm for $S_{\text{ToE}}$ to offer a complete and consistent description of reality.

 %\section{Axioms for a \(S_{\text{ToE}}\) Including Quantum Gravity}
Even though we do not have a  $S_{\text{ToE}}$, which explains all the fundamental forces, and produces a well-defined theory of quantum spacetime (quantum gravity), we can suggest some general axioms that  $S_{\text{ToE}}$ should satisfy.  
In constructing $S_{\text{ToE}}$ with quantum gravity taken into account, we propose the following axioms:

\bigskip
\noindent {\textit{Axiom 1. Self-Consistency: }}The $S_{\text{ToE}}$ is internally consistent. Thus,  where applicable, it is renormalizable. This shows that physical quantities remain finite and well-defined when computed to all orders in perturbation theory.

\bigskip
\noindent {\textit{Axiom 2. Completeness:}}
The $S_{\text{ToE}}$ by definition should explain every physical phenomenon in the universe/multiverse, and all physics should be derivable from it. Consequently, a $S_{\text{ToE}}$ must be inherently complete, i.e., it should encompass all fundamental laws and principles that govern the behavior in the universe/multiverse.

\bigskip
\noindent {\textit{Axiom 3. Fundamental Fields on a Configuration Space:}}
Now we should have a configuration space \(\chi\), where all quantum fields, and quantum spacetime metric, should have their degrees of freedom encoded. The fundamental basic objects in the theory should be fields \(\Psi(\chi)\) expressed on this configuration space \(\chi\).  For example in string theory, \(\chi\) describes string configurations and $\Psi (\chi)$ would describe a string field \cite{a1}, while for loop quantum gravity, \(\chi\) may be described by spin networks, and $\Psi(\chi)$ a  group fields i.e., field in third quantized loop quantum gravity
\begin{equation}
\Psi: \chi \rightarrow \mathbb{C}
\end{equation}

\bigskip
\noindent {\textit{Axiom 4. Quantum State Space and Hilbert Space Structure:}}
The quantum states of the theory form a Hilbert space \(\mathcal{H}\), provided with an inner product \(\langle \Psi, \Phi \rangle\) that permits the computation of probabilities and expectation values 
\begin{equation}
  |\langle \Psi, \Phi \rangle| \quad < \infty
\end{equation}
The form of the inner product will depend on the details of  $S_{\text{ToE}}$. 

\bigskip
\noindent {\textit{Axiom 5. Dynamics and Evolution - A Universal Action:}}
The dynamics of the fields are prescribed by an action \( S[\Psi] \), where the action is a functional of the fields.
\begin{equation}
S[\Psi]=\int_{\chi} d\mu(\chi) \left(\frac{1}{2}\langle \Psi, \mathcal{K} \Psi\rangle + \lambda \mathcal{V}(\Psi)\right)
\end{equation}
where \(d\mu(\chi)\) is a measure on the configuration space, \(\mathcal{K}\) kinetic operator and \(\mathcal{V}(\Psi)\) interaction potential with coupling constant \(\lambda\). For example, in string field theory  $\mathcal{K}$ will be the BRST operator \cite{a1}, and in group field theory $\mathcal{K}$ will be the appropriate Kinetic operator  \cite{a2}.

\bigskip
\noindent {\textit{Axiom 6. Gauge Symmetry:}}
The theory is invariant in the group of local gauge transformations \(G\), whose fields \(\Psi(\chi)\) transform under this group representation in which the action \(S[\Psi]\) remains invariant
\begin{equation}
\begin{aligned}
\Psi(\chi) \rightarrow \Psi^G(\chi)& = U(G)\Psi(\chi), \\
S[\Psi] &= S[\Psi^G]
\end{aligned}
\end{equation}
The details of this would depend on the exact nature of the theory.   

\bigskip
\noindent {\textit{Axiom 7.  Emergent Spacetime:}}
The spacetime and its geometry emerge from the quantum states of the gravitational field \cite{Faizal,Faiz1}. The metric tensor $g_{\mu \nu}$ and other geometric quantities are derived as effective fields from the quantum
state of the gravitational field and are only defined in appropriate approximations. 

\bigskip
\noindent {\textit{Axiom 8. Unification of Forces:}}
All forces, the gravitational, electromagnetic, weak, and strong forces, are unified into one framework \cite{u1, u2}. Different modes or different configurations of fundamental fields $\chi$ cause the appearance of different forces.

\bigskip
\noindent {\textit{Axiom 9. Quantization:}}
The theory respects the principles of quantum mechanics, where certain observables are quantized. It should give a natural internally consistent interpretation of quantum mechanics. 
The theory implies that the spectra of certain operators are discrete,
$\hat{Q} \Psi = \Lambda \Psi, \quad \Lambda \in \text{Discrete Set}$. These would correspond to fundamental constants of nature. Thus, fundamental constants of nature should also be obtained as consequences of this theory. 

\bigskip
\noindent {\textit{Axiom 10. Holographic Principle:}}
In the limit, in which a geometry emerges, the holographic principle should hold. 
The theory satisfies the holographic principle, i.e., the information content of a region of space shall be bounded by the degrees of freedom of that region's boundary \cite{holog},
\begin{equation}
\text{Entropy}(V) = \frac{\text{Area}(\partial V)}{4G}
\end{equation}
 
 %\section{Application of Tarski's Undefinability Theorem}
A major obstruction in the construction of a complete and consistent $S_{\text{ToE}}$, where all scientific truths are internally defined is the Tarski's undefinability theorem.
Under Tarski's undefinability theorem, any kind of formal system containing the axioms of the $S_{\text{ToE}}$  has some intrinsic limitations, i.e., the theorem states that any sufficiently rich formal system cannot describe or even define truth in itself. Thus, scientific truth cannot be defined internally within $S_{\text{ToE}}$.  Here, the formal system $S_{\text{ToE}}$ is definitely robust enough to cover the arithmetic. In that case, Tarski's undefinability theorem states that one cannot form a formula $T(x)$ which—within such a system—obeys the following:
\begin{equation}
\forall \phi \in S_{\textit{ToE}}, \quad S_{\textit{ToE}} \vdash T(\ulcorner \phi \urcorner) \leftrightarrow \text{True}(\phi)    
\end{equation}
 Thus, no formula can be defined inside the system ‐ $S_{\text{ToE}}$ that acts as a universal truth predicate — being correct in the determination of the truth of every statement ‐ under ‐ $S_{\text{ToE}}$.
In this context, $ \ulcorner \phi \urcorner $ represents the Gödel number of the formula $\phi$, which encodes the statement as a natural number. Such fine encoding allows the system $S_{\text{ToE}}$ to handle problems about $\phi$ as numerical problems. Using the diagonalization technique, very similar to that used by Gödel when proving his incompleteness theorems, one may derive a statement $G(x)$, which roughly speaking states its unprovability: $\>[ G(x) \equiv
\neg T(x)$
where \( T(x) \) is the assumed truth predicate.

Now let \( x = \ulcorner G \urcorner \), which is the Gödel number of \( G(x) \). Then the statement \( G(\ulcorner G \urcorner) \) says:
\begin{equation}
G(\ulcorner G \urcorner) \equiv \neg T(\ulcorner G \urcorner)\end{equation}
Now, if $T(\ulcorner G\urcorner)$ were true, then $G(\ulcorner G\urcorner)$ would be false. Yet this contradicts our assumption that $T(\ulcorner G\urcorner)$ is true. Next, if $T(\ulcorner G\urcorner)$ were false, then $G(\ulcorner G\urcorner)$ would be true, again giving rise to a contradiction.

The paradox then, proves that no universal truth predicate $T(x)$ can be written in the system $S_{\text{ToE}}$, thus proving Tarski's undefinability theorem. So,  for the formal system representing the axioms of the \(S_{\text{ToE}}\), we can write:
$
\neg \exists \quad T_{\text{$S_{\text{ToE}}$}}(x) \in S_{\text{$S_{\text{ToE}}$}} $   such that  $\forall \phi \in S_{\text{$S_{\text{ToE}}$}}, \quad S_{\text{$S_{\text{ToE}}$}} \vdash T_{\text{$S_{\text{ToE}}$}}(\ulcorner \phi \urcorner) \leftrightarrow \text{True}(\phi)
$.
In other words, even in $S_{\text{ToE}}$, which in principle encodes all the fundamental laws producing the universe/multiverse, it is not possible to formulate a single expression, $T_{{S_{\textit{ToE}}}}(x)$, that would be able to determine the truth of every conceivable statement in the theory. Among them are statements regarding quantum fields and spacetime, very basic structures that occur in the universe/multiverse.

The implications of this result are profound and vast. We start from all basic laws of physics, which produce the universe/multiverse, and define them as a formal system. Then, necessarily, this system would be incomplete concerning its truth. It follows that $S_{\text{ToE}}$ somehow must exit self-containment to reach some truth concerning itself beyond the formal system.

 %\section{Overcoming the Difficulty with the Lucas-Penrose Argument}

The Lucas-Penrose argument is an argument based on the observation that non-algorithmic understanding exceeds the threshold of restrictions held by formal systems such as $S_{\text{ToE}}$ \cite{penrose1989emperor, lucas1961minds}. The argument is based on Gödel's incompleteness theorems and argues that human minds can recognize truths that are not accessible to formal systems using algorithms due to their non-algorithmic understanding.

A Gödel sentence \(G_{\text{$S_{\text{ToE}}$}}\) in $S_{\text{ToE}}$ is a statement that asserts its own unprovability:
\begin{equation}
G_{\text{$S_{\text{ToE}}$}} \equiv \neg \text{Provable}_{S_{\text{$S_{\text{ToE}}$}}}(\ulcorner G_{\text{$S_{\text{ToE}}$}} \urcorner)
\end{equation}
Where \(\text{Provable}_{S_{\text{$S_{\text{ToE}}$}}}(x)\) is the formal predicate indicating that \(x\) is provable within $S_{\text{ToE}}$. The Gödel sentence \(G_{\text{$S_{\text{ToE}}$}}\) has the following properties:
\begin{equation}
S_{\text{$S_{\text{ToE}}$}} \nvdash G_{\text{$S_{\text{ToE}}$}} \quad \text{and} \quad S_{\text{$S_{\text{ToE}}$}} \nvdash \neg G_{\text{$S_{\text{ToE}}$}} 
\end{equation}
This means that \(G_{\text{$S_{\text{ToE}}$}}\) is neither provable nor disprovable within $S_{\text{ToE}}$. However, \(G_{\text{$S_{\text{ToE}}$}}\) is true because if it were false, then it would be provable, leading to a contradiction.

The Lucas-Penrose argument asserts that a human mind due to non-algorithmic understanding, and reasoning outside the formal system $S_{\text{ToE}}$, can recognize the truth of \(G_{\text{$S_{\text{ToE}}$}}\). The reasoning follows:
A human can make sense that if \(G_{\text{$S_{\text{ToE}}$}}\) were false, then it would make $S_{\text{ToE}}$ true, thus deriving the system into inconsistency.
Thus, the human mind due to non-algorithmic understanding can assert the truth of \(G_{\text{$S_{\text{ToE}}$}}\) without relying on the formal system.
Knowing \(G_{\text{$S_{\text{ToE}}$}}\) implies that there is something about the truth that goes beyond the computational algorithms, i.e., no computational algorithmic process in $S_{\text{ToE}}$ can fully capture the truth of \(G_{\text{$S_{\text{ToE}}$}}\).

The Lucas–Penrose argument proposes a way for $S_{\text{ToE}}$ to encompass the full formal structure of everything. This is done using non-algorithmic understanding, which overcomes the inherent limitations of computational algorithms. In particular, by recognizing the truth of certain statements, which the computational algorithms are not able to prove,  non-algorithmic understanding can go beyond the limitations of  Tarski's undefinability theorem for any formal system—hence, also of $S_{\text{ToE}}$. So, to fully understand $S_{\text{ToE}}$, requires non-algorithmic understanding. 
However, humans understand $S_{\text{ToE}}$,  but 
$S_{\text{ToE}}$ actually exists in a Platonic realm. It is important to distinguish between these two concepts, i.e. human understanding of $S_{\text{ToE}}$ being consistent and  $S_{\text{ToE}}$ being consistent (not human understanding of  $S_{\text{ToE}}$, but the real  $S_{\text{ToE}}$, which exists in the actual  Platonic realm).  It is possible for human understanding to be inconsistent, and hence the human understanding of  $S_{\text{ToE}}$ to be inconsistent, and the criticism of the Lucas-Penrose argument is based on the assumption that human thought is inconsistent \cite{d1}.  Even though there is a debate on this matter \cite{Penrose1996}, a disjunction holds for human thought, i.e. human thought is either non-algorithmic or inconsistent.  However, these criticisms do not apply to the actual $S_{\text{ToE}}$ in the  Platonic realm, as it is impossible for actual $S_{\text{ToE}}$ in the  Platonic realm to be inconsistent, as this would produce real inconsistencies in the universe/multiverse. 
Now for truth in the real  $S_{\text{ToE}}$, in the Platonic realm, to be internally defined (which is the bare minimum requirement of $S_{\text{ToE}}$), a non-algorithmic understanding has to operate in the Platonic realm, apart from computational algorithms converting the consequences of  $S_{\text{ToE}}$ into reality, and producing the universe/multiverse and effective physical laws in it.  Thus, since it is impossible to use algorithms to obtain self-consistency and completeness for $S_{\text{ToE}}$, the ultimate theory about reality has to have truths in it, that can only be acquired by a non-algorithmic understanding operating in the Platonic realm. These truths are also accessible to humans because the human mind can also grasp them through non-algorithmic understanding. However, the very notion of truths exists in $S_{\text{ToE}}$, and hence a non-algorithmic understanding also exists within the Platonic realm producing the complete and consistent physics in the universe/multiverse.  
 
% \section{Conclusion}  
In conclusion, a \(S_{\text{ToE}}\)  aims to unify all fundamental forces of nature, including quantum gravity, into a single, coherent theoretical framework. Such a theory would be internally defined by a set of axioms, from which all scientific truths could be derived as consequences. This paper has proposed a set of axioms for any potential $S_{\text{ToE}}$ and explains how scientific truth would be viewed as a consequence of $S_{\text{ToE}}$. We then explored the implications of Tarski's undefinability theorem on the scientific truths that can be derived from these axioms. We have demonstrated that Tarski's theorem imposes inherent limitations on any formal system attempting to fully encapsulate a $S_{\text{ToE}}$. However, we also argue that the Lucas-Penrose argument demonstrates that non-algorithmic forms of understanding might transcend these formal limitations. 

 Thus, any unified framework in physics that can provide a complete and consistent description of reality must be produced by non-algorithmic understanding in the Platonic realm. So,  no $S_{\text{ToE}}$, formulated as a formal system with computational algorithms, could ever internally define the truth. Tarski's theorem thus shows it would be impossible to specify within the system a general truth predicate; or, more precisely,     $S_{\text{ToE}}$ will be restricted in its capacity for self-validation and full truth-containment. The Lucas-Penrose argument surpasses this limitation. So, to surpass the limitations of such a formal system $S_{\text{ToE}}$ requires non-algorithmic understanding. Human non-algorithmic understanding testifies to this, being able to see truths that are completely beyond algorithmic operations.  We also argue that such non-algorithmic understanding exists not only in the human mind but also in the Platonic realm, which produces the universe/multiverse with effective laws of physics in it. 
 This is the only way for the actual  $S_{\text{ToE}}$ (not human understanding of  $S_{\text{ToE}}$) to overcome the limitations of Tarski's theorem. 
 Ultimately, the definition of scientific truth in the universe/multiverse lies beyond any form of computational construction, as it rests in the non-algorithmic understanding in the Platonic realm. 

\bibliographystyle{ieeetr}

\begin{thebibliography}{99}

\bibitem{Galilei1967}
G.~Galilei, \emph{Dialogue Concerning the Two Chief World Systems}, 2nd ed., University of California Press, 1967.

\bibitem{Popper1959}
K.~Popper, \emph{The Logic of Scientific Discovery}, Hutchinson, London, 1959.

\bibitem{Weinberg1994}
S.~Weinberg, \emph{Dreams of a Final Theory: The Scientist's Search for the Ultimate Laws of Nature}, Vintage, 1994.
 
\bibitem{bl} M.~Maggiore, A generalized uncertainty principle in quantum gravity,  {Phys. Lett. B}  {304} (1993) 65-69.

\bibitem{le} I.~Pikovski, M.~R.~Vanner, M.~Aspelmeyer, M.~Kim and C.~Brukner, Probing Planck-scale physics with quantum optics,  {Nature Phys.}  {8} (2012) 393.

\bibitem{ti} M.~Faizal, M.~M.~Khalil and S.~Das, Time measurement uncertainty relations from generalized uncertainty principle,  {Eur. Phys. J. C}  {76} (2016) 30.

\bibitem{cut} A.~Kempf, Covariant information-density cutoff in curved space-time,  {Phys. Rev. Lett.}  {103} (2009) 231301.

\bibitem{pert} D.~Amati, M.~Ciafaloni and G.~Veneziano, Can spacetime be probed below the string size?,  {Phys. Lett. B}  {216} (1989) 41-47.

\bibitem{tdual} M.~Fontanini, E.~Spallucci and T.~Padmanabhan, Zero-point length from string fluctuations: The proof,  {Phys. Lett. B}  {633} (2006) 627-630.

\bibitem{c1} A.~Belenchia, D.~M.~T.~Benincasa and F.~Dowker, Quantum inhomogeneities in causal set cosmology,  {Class. Quant. Grav.}  {33} (2016) 245018.

\bibitem{c2} S.~A.~Wilkinson and A.~D.~Greentree, Causal dynamical triangulations with a single spatial slice,  {Phys. Rev. D}  {90} (2014) 124003.

\bibitem{c4} J.~Ambjorn, D.~Coumbe, J.~Gizbert-Studnicki and J.~Jurkiewicz, Scaling in four-dimensional causal dynamical triangulations,  {Phys. Rev. D}  {93} (2016) 104032.

\bibitem{nc10} M.~Rinaldi, Noncommutative geometry and the short-scale structure of spacetime,  {Class. Quant. Grav.}  {28} (2011) 105022.

\bibitem{nl10} R.~Brout, C.~Gabriel, M.~Lubo and P.~Spindel, Nonlocality in string-inspired field theory,  {Phys. Rev. D}  {59} (1999) 044005.

\bibitem{c5} A.~Ashtekar and P.~Singh, Loop quantum cosmology: A status report,  {Class. Quant. Grav.}  {28} (2011) 213001.

{\bibitem{c51} M.~Bojowald, Quantum cosmology: Effective theory,  {Nature Phys.}  {3} (2007) 523-525.}

\bibitem{c6} A.~Ashtekar and E.~Bianchi, A short review of loop quantum gravity,  {Rept. Prog. Phys.}  {84} (2021) 042001.

\bibitem{c10} A.~Perez, Spin foam models for quantum gravity,  {Class. Quant. Grav.}  {20} (2003) R43.

\bibitem{ph1} S.~W.~Hawking and R.~Penrose, The singularities of gravitational collapse and cosmology,  {Proc. Roy. Soc. Lond. A}  {314} (1970) 529.

\bibitem{ph2} R.~Penrose, Gravitational collapse and space-time singularities,  {Phys. Rev. Lett.}  {14} (1965) 57.

\bibitem{li1} S.~Alsaleh, L.~Alasfar, M.~Faizal and A.~F.~Ali, Lorentz violations in nonlocal quantum gravity,  {Int. J. Mod. Phys. A}  {33} (2018) 1850052.

\bibitem{li2} E.~C.~Vagenas, L.~Alasfar, S.~M.~Alsaleh and A.~F.~Ali, Noncommutative geometry and generalized uncertainty principle,  {Nucl. Phys. B}  {931} (2018) 72-78.

\bibitem{tdual1} H.~Wu and H.~Yang, Aspects of holography in higher dimensions,  {JCAP}  {07} (2014) 024.

\bibitem{lqgs} G.~Date and G.~M.~Hossain, Genericity of big bounce in isotropic loop quantum cosmology,  {Phys. Rev. Lett.}  {94} (2005) 011302.

\bibitem{gup} M.~Salah, F.~Hammad, M.~Faizal and A.~F.~Ali, Spherical symmetry and higher order generalized uncertainty principle,  {JCAP}  {02} (2017) 035.

\bibitem{gup2} A.~Awad and A.~F.~Ali, Minimal length and quantum gravity effects in the early universe,  {JHEP}  {1406} (2014) 093.

\bibitem{jc} T.~Jacobson, Thermodynamics of spacetime: The Einstein equation of state,  {Phys. Rev. Lett.}  {75} (1995) 1260.

\bibitem{hpbe} R.~Bousso and A.~Shahbazi-Moghaddam, Consistency of the entropy bound and the generalized second law,  {Phys. Rev. Lett.}  {128} (2022) 231301.


\bibitem{holog} R.~Bousso, The holographic principle,  {Rev. Mod. Phys.}  {74} (2002) 825-874.

\bibitem{ads} J.~L.~Petersen, Introduction to the Maldacena conjecture on AdS/CFT,  {Int. J. Mod. Phys. A}  {14} (1999) 3597-3672.


 

\bibitem{Faizal}M.~Faizal,
The end of spacetime,
Int. J. Mod. Phys. A {38}   (2023) 2350188.

 


\bibitem{penrose1989emperor} R.~Penrose, The Emperor's New Mind: Concerning Computers, Minds, and the Laws of Physics, Oxford University Press, 1989.

\bibitem{lucas1961minds} J.~R.~Lucas, Minds, Machines and Gödel,  {Philosophy} 



\bibitem{t1}A. Tarski, The Concept of Truth in Formalized Languages, in Logic, Semantics, Metamathematics: Papers from 1923 to 1938, J. H. Woodger trans., Oxford: Clarendon Press (1956).

\bibitem{t2} A. Tarski, Der Wahrheitsbegriff in den formalisierten Sprachen, Studia Philosophica 1 (1933) 261-405.

\bibitem{a1} E. Witten, Noncommutative Geometry and String Field Theory, Nucl. Phys. B 268 (1986) 253-294.
\bibitem{a2} A.~Baratin and D.~Oriti, Group field theory with non-commutative metric variables, Phys. Rev. Lett. 105 (2010) 221302.

\bibitem{Faiz1}S.~L.~Braunstein, M.~Faizal, L.~M.~Krauss, F.~Marino and N.~A.~Shah, Analogue simulations of quantum gravity with fluids,
Nature Rev. Phys.  {5} (2023) 612. 
%\bibitem{Faiz2}M.~Faizal,
%The end of spacetime,
%Int. J. Mod. Phys. A {38}   (2023) 2350188. 
\bibitem{u1}A.~Salam, Gauge Unification of Fundamental Forces,
Rev. Mod. Phys. {52} (1980) 525.
\bibitem{u2} M.~B.~Green,
Unification of forces and particles in superstring theories, 
Nature {314}  (1985) 409. 

\bibitem{d1} M.~Minsky, Conscious Machines, Machinery of Consciousness, Proceedings, National Research Council of Canada, 75th Anniversary Symposium on Science in Society,  (1991).
\bibitem{Penrose1996} R.~Penrose, Beyond the Doubting of a Shadow, Psyche  (1996) 23.


\end{thebibliography}

\end{document}